\documentclass[apj]{emulateapj}

\shorttitle{IR anisotropy in powerful AGN}
\shortauthors{H\"onig et al.}

\begin{document}


\title{Quantifying the anisotropy in the infrared emission of powerful AGN}


\author{S.~F. H\"onig,\altaffilmark{1} C. Leipski,\altaffilmark{2} R. Antonucci,\altaffilmark{1} M. Haas\altaffilmark{3}}
\altaffiltext{1}{University of California in Santa Barbara, Department of Physics, Broida Hall, Santa Barbara, CA 93109, USA; shoenig@physics.ucsb.edu}
\altaffiltext{2}{Max-Planck-Institut f\"ur Astronomie, K\"onigstuhl 17, 69117 Heidelberg, Germany}
\altaffiltext{3}{Astronomisches Institut Ruhr–Universit\"at Bochum, Universit\"atsstr. 150, 44801 Bochum, Germany}




\begin{abstract}
We use restframe near- and mid-IR data of an isotropically selected sample of quasars and radio galaxies at $1.0 \le z \le 1.4$, which have been published previously, to study the wavelength-dependent anisotropy of the IR emission. For that we build average SEDs of the quasar subsample (= type 1 AGN) and radio galaxies (= type 2 AGN) from $\sim1-17\,\micron$ and plot the ratio of both average samples. From 2 to 8\,$\micron$ restframe wavelength the ratio gradually decreases from 20 to 2 with values around 3 in the 10\,$\micron$ silicate feature. Longward of 12\,$\micron$ the ratio decreases further and shows some high degree of isotropy at 15\,$\micron$ (ratio $\sim$1.4). The results are consistent with upper limits derived from the X-ray/mid-IR correlation of local Seyfert galaxies. We find that the anisotropy in our high-luminosity radio-loud sample is smaller than in radio-quiet lower-luminosity AGN which may be interpreted in the framework of a receding torus model with luminosity-dependent obscuration properties. It is also shown that the relatively small degree of anisotropy is consistent with clumpy torus models.
\end{abstract}





\section{Introduction}

In the unification scheme of AGN the difference between type 1 and type 2 AGN is explained by angle-dependent circumnuclear obscuration of the accretion disk and broad-line region \citep{Ant93}. This obscuring dusty medium -- commonly referred to as ``dust torus'' -- is optically and geometrically thick and probably extends from sub-parsec scales outward to several 10s of parsecs, or beyond for high luminosity objects. The dust in the torus absorbs the incident UV/optical radiation an re-emits the received energy in the infrared. 

Observations have shown that type 1 AGN show significantly more emission in the near-IR than type 2 AGN for the same given intrinsic luminosity \citep[e.g.][]{Lei05}. This is consistent with the picture where the face-on view onto the torus in type 1 AGN exposes the innermost hot dust to the observer. On the other hand in type 2 AGN the torus is seen edge-on so that internal obscuration blocks the line-of-sight to the hot dust. Owing to this effect, it is expected that for a given AGN luminosity the infrared emission of type 1 AGN is generally stronger than from type 2s. 

In the light of attempts at forming isotropic AGN samples based on IR fluxes it seems important to know exactly how strong of a bias towards type 1s over type 2s may occur when invoking flux limits. Moreover, probing the wavelength dependence of this anisotropy in the infrared has some constraining power on our understanding of how the torus obscures the AGN. It may be possible to distinguish torus models where the dust is smoothly distributed from those where the dust is arranged in clouds \citep[e.g.][]{Lev09}: If the dust is smoothly distributed within the torus, a large degree of anisotropy is expected. If, however, the dust is arranged in clouds the anisotropy is expected to be smaller.

A problem commonly encountered when studying AGN samples in the local universe is a significant contribution of the host galaxy to the IR. This is related to the typical lower luminosity Seyfert galaxies which dominate the nearby AGN population. One way around this problem is the use of high-spatial resolution observations, as possible with the largest ground-based telescopes or interferometers, which are able to resolve out the host and isolate the AGN emission \citep[for details see][]{Hon10}. However, it is difficult to set up representative samples owing to the observational limitations. Another possibility is the use of high luminosity objects -- typically at higher redshift -- where the AGN outshines the host galaxy by a large factor in the optical and near-IR. If PAH features are absent, the AGN most likely dominates the mid-IR wavelength region as well (in our sample: host $\ll 10\%$ for wavelengths $\le$ 17\,$\micron$).

In this paper we aim at quantifying the wavelength dependence of the anisotropy of the AGN emission in the infrared from $\sim1-17\,\micron$. For that we use a (nearly) isotropically selected and complete sample of quasars and radio galaxies with hidden quasars at $1.0 \le z \le 1.4$ as recently presented in \citet{Lei10} and described in Sect.~\ref{sec:sample}. Here we improve the analysis by using host-galaxy subtraction for the radio galaxies and use clumpy torus models for interpretation. In Sect.~\ref{sec:results} we show the average SEDs of each of the subsamples which are representing obscured (type 2) and unobscured (type 1) AGN. We further analyze the origin of the anisotropy by fitting extinction models and clumpy torus models to the observations in Sect.~\ref{sec:analysis}. In Sect.~\ref{sec:obsaniso} we discuss our results by comparing them to previous IR anisotropy estimates in literature. The results are summarized in Sect.~\ref{sec:summary}.

\section{Sample selection and data}\label{sec:sample}

The object sample for this paper comprises all 3CRR \citep{Lai83}
radio galaxies and quasars with $1.0 < z < 1.4$
This lobe dominated sample presents a well
matched set of radio galaxies and quasars in terms of their intrinsic
luminosity  ($\langle\nu{\rm L}_{\nu,178\,{\rm MHz}}\rangle = (1.9 \pm
0.6) \times 10^{44}$\,erg/s for the quasars and $\langle\nu{\rm
L}_{\nu,178\,{\rm MHz}}\rangle = (2.0 \pm 0.4) \times 10^{44}$\,erg/s
for the radio galaxies; errors indicate standard deviation of the sample). The data used here have been presented
previously in \citet{Haa08} and \citet{Lei10} and we refer the reader
to these papers for further details on the source  selection, data
reduction, and building of the average SEDs. To summarize  briefly, we
obtained mid-IR photometry in all six filters from 3.6\,$\micron$ to
24\,$\micron$ and spectroscopy from 19\,$\micron$ to 38\,$\micron$
utilizing all three instruments onboard the {\it Spitzer} Space
Telescope \citep{Wer04}. After the data reduction in a standard
manner, the  individual source SEDs were interpolated onto a common
rest frame wavelength grid. The quasars and  radio galaxies were then
averaged 
into a mean SED for each class of objects. The
individual SEDs (including observed and interpolated photometry) as
well as the the average SEDs are presented in
\citet[][their Figs.\,1 \& 2]{Lei10}.

For this paper, additional corrections have been applied to the radio
galaxy data before the averaging process outlined above: At the
shortest wavelengths considered here ($\sim 1-3$\,$\mu$m, rest frame)
the radio galaxy SEDs show contributions from the host galaxy. Since
we want to isolate the emission coming from the active nucleus, we
have to correct for the stellar emission in these cases.  This
correction was performed by fitting the observed IRAC photometry with
a combination of a moderately old (5-10\,Gyr)
elliptical galaxy SED to represent the stellar emission (taken from
the GRASIL webpage; \citealt{Sil98}) and a hot black body for the AGN
dust whose temperature we allowed to vary 
\citep[see e.g.][]{Sey07,Deb10}. The resulting black body temperatures 
in the radio galaxies range from 600 to 970\,K, with a median value of 860\,K. The fraction of host galaxy light contributing to 
the flux measured at the observed frame wavelengths 3.6, 4.5, 5.8, and 
8.0\,$\mu$m was found to be 0.9, 0.6, 0.3, and 0.1, respectively.
Despite a negligible contribution from the host galaxy, we performed similar fits to the quasar sample as well to obtain characteristics of the hot dust emission and compare it to the radio galaxies. For the quasars we find notably hotter temperatures in the range from 880--1250\,K (median 1020\,K). This is consistent with the idea that the hottest dust is obscured in radio galaxies, while directly seen in quasars.

In the radio galaxies we then subtracted the estimated host
galaxy contribution from the
observed IRAC photometry using the scaled template SED.  
For the IRS and MIPS measurements the host galaxy contributions were
considered negligible at restframe wavelength $\gtrsim 7\,\micron$ 
and no corrections have been applied. We refrain from further corrections related to possible starformation. As pointed out in \citet{Lei10} neither the individual SEDs nor the averages showed any PAH features (see also Fig.~\ref{fig:aver_sed}), which indicates that any contribution from starformation to the mid-IR is probably negligible.

The radio galaxy
3C\,368 has a galactic M-star superimposed close to the position of
the radio galaxy nucleus \citep[e.g.][]{Ham91,Bes97}. Both sources
are partly blended  even at the shortest IRAC wavelengths which makes
the correction for the  host galaxy emission in this source quite
uncertain. Consequently, we removed  this source from the sample
considered here. This leaves us with 11 quasars and 8 radio galaxies
from which the average SEDs have been calculated.

\section{Infrared SED of quasars and radio galaxies at $z\sim1.2$}\label{sec:results}

\begin{figure}
\epsscale{1.2}
\plotone{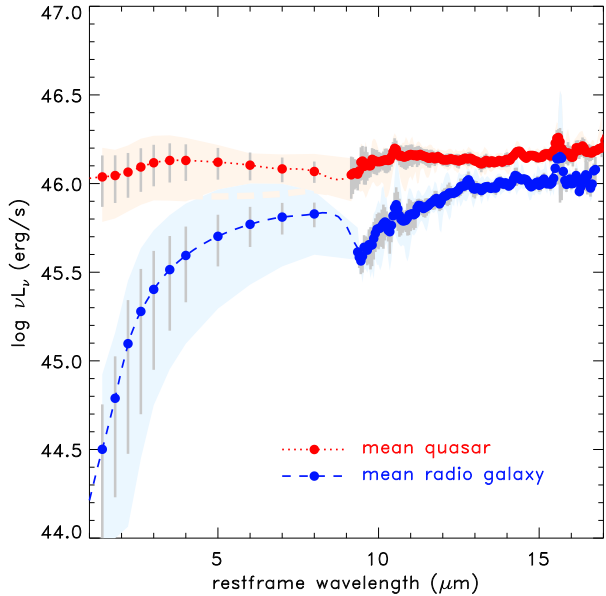}
\caption{Infrared sample average SED of quasars (red filled circles) and radio galaxies (blue filled circles). The gray error bars represent the dispersion (mean absolute deviation) of each sample while the red- and blue-shaded areas indicate the total range covered by the quasar and radio galaxy sub-samples, respectively. The red-dotted and blue-dashed lines are cubic spline fits to the quasar and radio galaxy average SED. \label{fig:aver_sed}}
\end{figure}

In Fig.~\ref{fig:aver_sed} we show the average SED of $1.0<z<1.4$ quasars (red) and radio galaxies (blue) respectively. The error bars reflect the mean absolute deviation while the shaded areas show the range of the respective subsample at each wavelength point of the interpolated data (see Sect.~\ref{sec:sample}). For each of the object types we calculated a spline fit through the mean data points in order to guide the eye. It is obvious that the radio galaxies are systematically lower in infrared emission than the quasars. The discrepancy is largest in the near-IR and flattens out towards longer wavelengths. \citet{Lei10} showed similar average SEDs. Here we used additional host galaxy subtraction, which isolates the AGN light much better (see Sect.~\ref{sec:sample}). This is most obvious in the near-IR part of the radio galaxies shortward of 5\,$\micron$. The SED keeps on falling toward shorter wavelengths consistent with the Wien tail from hot dust emission, instead of making an upward turn \citep[see][Fig. 2]{Lei10}.

Since both types have the same radio luminosities due to our selection, this difference between quasars (= type 1 AGN)  and radio galaxies (= type 2 AGN) is a generic property of the sample. It either reflects a difference in line-of-sight extinction (e.g. by cold dust in the host galaxy), or traces the anisotropy in re-emission of the AGN-heated dust. \citet{Lei10} tested the former possibility and found a surprisingly good match of the difference between radio galaxies and quasars in the mid-IR by a single extinction law. This, however, breaks down in the near-IR. In this paper we will test if, instead, a single absorber and emitter (= the dust torus) may be responsible for the radio galaxy/quasar anisotropy (see Sect.~\ref{sec:modcomp}).

\begin{figure}
\epsscale{1.2}
\plotone{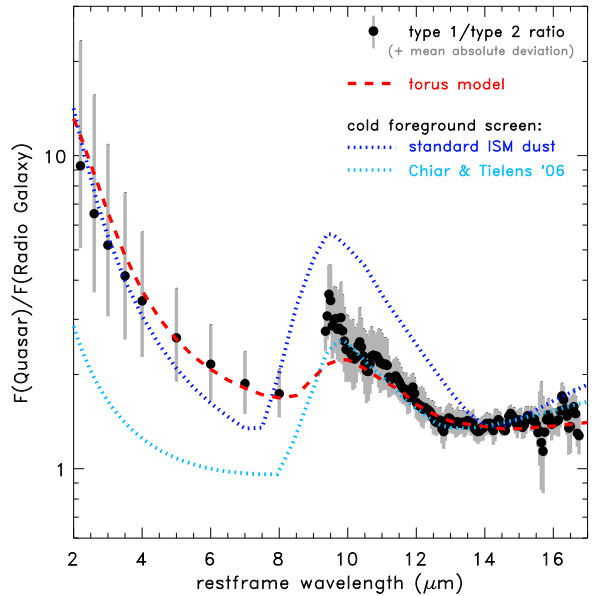}
\caption{Wavelength dependence of the ratio of quasar to radio galaxy SED as a measure of anisotropy of the infrared emission. The black-filled circles are the ratios of the average SEDs as shown in Fig.~\ref{fig:aver_sed} and the dark-gray error bars are the mean absolute deviation for each wavelength bin. The red-dashed line is a comparison to clumpy torus models. The dark-blue dotted line shows a standard ISM dust screen extinction curve scaled to the type 1/type 2 ratio at 15\,$\micron$, while the light-blue dotted lines is screen extinction using the IR absorption curve by \citet{Chi06}. For details see text.} \label{fig:ratio}
\end{figure}

To quantify the wavelength dependence of the anisotropy we plot the quasar/radio galaxy ratio in Fig.~\ref{fig:ratio}. Also shown is the mean absolute deviation of the ratio calculated by propagating the standard deviations of each subsample. From 2 to 8\,$\micron$ the emission ratio gradually decreases from 20 to 2. In the silicate feature this ratio increases again up to about 3 and flattens out towards 15\,$\micron$ at a value of $1.44\pm0.17$. 

\begin{figure}
\epsscale{1.2}
\plotone{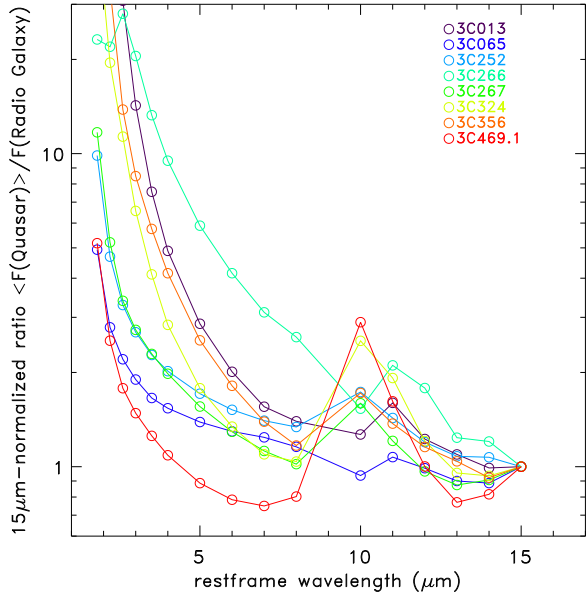}
\caption{Wavelength dependence of the ratio for each individual radio galaxy using the mean quasar spectrum $<\!\!F(\mathrm{Quasar})\!\!>$. All individual ratios have been normalized to 15\,$\micron$ (and the IRS range $>10\,\micron$ has been binned) for illustrating the range covered by our sample.} \label{fig:range}
\end{figure}

Our sample comprises a range of SED shapes, meaning that a range of ratios is observed. Most of this sample range comes from the radio galaxies which show much less uniformity than the quasars (see Fig.~\ref{sec:sample}). We illustrated the range of ratios covered by the radio galaxies in Fig.~\ref{fig:range} where we plot the wavelength dependence of the ratio of each radio galaxy using the average quasar SED and normalize it to 15\,$\micron$.

At around 10\,$\micron$ the radio galaxies show the silicate feature in absorption while quasars display a weak silicate emission feature. We used the spline fits described above (see Fig.~\ref{fig:aver_sed}) to locate the centers of the silicate features following the method outlined by \citet{Sir08}. The silicate absorption feature center is found to be at $9.6\pm0.1\,\micron$ while the peak of the silicate emission feature was measured at $10.6\pm0.1\,\micron$. This difference in central wavelength is similar to the ones observed in local galaxies. As recent high spatial resolution studies of Seyferts suggest, the ``shift'' in wavelength is not a pure radiative transfer effect due to the location or distribution of the dust, but implies a change of dust chemistry within the torus \citep{Hon10}. As demonstrated by \citet{Smi10} the silicate emission feature may be located at around 10.5\,$\micron$ if a fraction of the hot silicate dust consists of porous silicate grains.

We note that the transition from quasars to radio galaxies is not as smooth as one may expect. In spite of some overlap in the range of SEDs of both types in the near-IR part in Fig.~\ref{fig:aver_sed}, quasars generally show infrared emission characteristics expected for a type 1 AGN (blue IR color; silicate emission feature) and radio galaxies have IR SEDs with type 2 emission properties (redder SED; silicate absorption feature).

\section{Analysis}\label{sec:analysis}

In this section we analyze the scenarios that can lead to the observed anisotropy in radio galaxies and quasars. There are two possibilities: (1) extinction by a cold dust screen and (2) absorption and emission in a warm dusty medium. The former possibility may be associated with cold dust in the host galaxy while the latter one is equivalent to the dust torus in the AGN unification scheme (which we may call ``intrinsic anisotropy'').

\subsection{Cold dust screen extinction}\label{sec:extinct}

We will first discuss the plausibility and consequences of a cold dust screen on the SEDs. If the IR anisotropy is dominated by cold host galaxy dust, then we have a situation where the IR emission originates from the torus while the absorption is coming from a different component (i.e. dust in the host). This means that we require an additional component outside the AGN to model the data (e.g. as used for a significant minority of high-$z$ type 2 QSOs in \citet{Pol08}). However, the objects suffering extinction (here: radio galaxies) would be offset from quasars by only a single extinction law. This has been tested and ruled out by \citet{Lei10} for the same set of objects as presented here. In Fig.~\ref{fig:ratio} we show the anisotropy ratio between quasars and radio galaxies (black circles with error bars). We overplot a standard ISM extinction curve, resembling cold screen extinction, scaled to the observed 15\,$\micron$ anisotropy (dark-blue dotted line; using a mixture of 53\% silicates and 47\% graphite, based on updated dust opacity cuves by \citet{Dra03}, and a grain size distribution according to \citet{Mat77}). The extinction curve significantly overpredicts the anistropy in the silicate feature with better agreement in the near-IR. For reference we also plot the extinction curve based on \citet[][light-blue dotted line; ``Pixie dust'']{Chi06}, as used in \citet{Lei10}, which results in better agreement within the silicate feature but significant offsets in the near-IR. In fact, \citet{Lei10} pointed out that a good correspondence of quasars and radio galaxies can be achieved only if the quasar SED is attenuated by at least \textit{two} instead of one extinction components -- which is reminiscent of radiative transfer (= absorption \textit{and} emission) within the torus rather than a cold screen. 

Moreover, we found a mean anisotropy of about 1.4 at 15\,$\micron$ (see Sect.~\ref{sec:results}). Dust opacity curves typically have opacity ratios $\tau_\mathrm{15\,\micron}/\tau_V \sim 0.01$. In order to obtain the observed anisotropy, the dust screen would have to have $\tau_V\sim30$. While such optical depth values are in reach for galactic dust lanes (e.g. extinction towards our own Galactic center), it requires very edge-on views onto disk galaxies since the scale height of galactic disks is small. On statistical grounds this possibility may be viable for a minority of all radio galaxies, but it is unlikely that the whole population is dominated by host extinction. We note that the same line of argument can be made using the near-IR anisotropy leading to even higher $\tau_V$ and illustrating the need for more then just one cold extinction screen in this scenario. 

\subsection{Torus model fitting}\label{sec:modcomp}

We use our clumpy torus models \textit{CAT3D} \citep{Hon10b} to test if the observed anisotropy ratio can be explained by the intrinsic anisotropy as predicted in the unification scheme. Generally smooth dust torus models predict stronger anisotropy than clumpy models \citep{Sch08}. \citet{Lev09} argue that the observed small anisotropy in the mid-IR/X-ray correlation (see Sect.~\ref{sec:obsaniso}) is qualitatively in agreement with a clumpy torus. Here we aim at being more quantitative and show consistency of the observed anisotropy with torus orientation.

In Fig.~\ref{fig:ratio} we overplot the observed ratio with predicted ratios of a clumpy torus model (red dashed line). As mean torus inclination in quasars we assumed 39$^\circ$ while the mean radio galaxy inclination is set to 75$^\circ$. This corresponds to a mean opening angle of the torus of 60$^\circ$ or a type 1/type 2 ratio of 1:1, which is consistent with the number statistics of our complete and isotropic sample \citep{Lei10}. Fig.~\ref{fig:ratio} shows that the model is following the continuum anisotropy curve quite well. There is, however, a slight deviation within the silicate feature where in the center of the feature the model curve predicts slightly lower anisotropy than shown by the observed curve. This may be indicative of additional, off-torus obscuration in some objects (e.g. from the host galaxy), which effectively deepens silicate absorption features in radio galaxies (the denominator of the plot in Fig.~\ref{fig:ratio}) and, to a lesser degree, changes the spectral slope. If this happens in individual objects, the average curve and scatter will tend to be slightly more anisotropic. Some local examples of host obscuration are Centaurus A, NGC~5506, or the nucleus of the Circinus galaxy where host-galactic dust lanes are projected onto the nucleus producing deep silicate features. In fact the silicate absorption feature in the average radio galaxy SED seems to be deeper than in typical local Seyfert 2 galaxies without host obscuration \citep{Hon10}. Note that this requires a statistical alignment of cold host dust with that of the torus (see Sect.~\ref{sec:extinct}).

The model used for reproducing the ratio uses a dust cloud distribution with radial power law  $\propto r^{-1.25}$ and 5 clouds along the line-of-sight in equatorial direction \citep[for details see][]{Hon10b}. In comparison to models for SEDs and mid-IR interferometry of local Seyfert galaxies, these parameters suggest only a slightly more centrally condensed and transparent torus \citep{Hon10}, while the half-opening angle may be wider (60$^\circ$ instead of 45$^\circ$). In fact a range of torus model parameters satisfies the observed type 1/type 2 anisotropy spectrum within the error bars of the sample (e.g. various steeper and shallower dust distributions). From Bayesian inference analysis we found that the torus model parameters are generally poorly constrained (broad posterior distributions for individual parameters). A weakness is certainly that modeling the flux ratio is not very constraining for model parameters since it does only take relative fluxes into account, while absolute fluxes (e.g. actual silicate strength of type 1s or type 2s) are not included. On the other hand, what the modeling shows is that the observed small ratios at long wavelengths and the change of anisotropy from the near- to the mid-IR are in agreement with expectations from clumpy torus models without fine-tuning parameters.

In summary, the torus model seems to reproduce the observed anisotropy reasonably well over most of the wavelength range, while single extinction laws result in much worse fits. The model parameters used in the clumpy torus model fit are reasonable in comparison to fits to local AGN. It demonstrates plausibility of the scenario that the anisotropy is related to torus orientation. This strongly suggests that the observed anisotropy is a measure for the intrinsic anisotropy of luminous type 1 and type 2 AGN at $z\sim1.2$.



\section{Other anisotropy estimates}\label{sec:obsaniso}

\subsection{Comparison to the mid-IR/X-ray correlation}

In Sect.~\ref{sec:analysis} we argued that the observed IR anisotropy is probably reflecting the ``intrinsic anisotropy'' as caused by the dust torus. The isotropic and complete selection of the sample helps to minimize any biasing effects on the anisotropy. On the other hand, optically- and X-ray-selected samples often suffer from missing some of the most obscured objects. Moreover, since the ratio is $>1$ at 15\,$\micron$, flux-limited mid-IR selected samples are potentially biased towards type 1 AGN as well. Thus, even using 12\,$\micron$-selected AGN samples will slightly suffer from the assumption of isotropy.

One popular way of comparing type 1s and type 2s is the correlation between X-ray and mid-IR luminosity. In case the X-luminosity is emitted isotropically (and traces the dust-heating emission), and the mid-IR emission is radiated very anisotropically, this correlation is expected to be different for type 1 and type 2 AGN. However, using high spatial resolution observations \citet{Gan09} showed that local Seyfert 1 and Seyfert 2 galaxies, up to column densities of few $10^{24}$ cm$^{-2}$, essentially follow the same correlation. A conservative estimate suggests that at 12\,$\micron$ the difference between both types is smaller than a factor of 3. Despite including some ``mildly'' Compton-thick objects in this study, the most obscured objects are still missing due to the lack of intrinsic X-ray data. This essentially makes anisotropy estimates from the mid-IR/X-ray-correlation a lower limit on the ``true'' anisotropy. Nevertheless this result is fully consistent with our finding for powerful $z\sim1.2$ AGN. At 12\,$\micron$ we find a type 1/type 2 ratio of $1.65\pm0.26$. Assuming that X-ray selection misses the highest-inclination objects our result would predict that the anisotropy in the mid-IR/X-ray correlation is $<1.6$, at least for powerful AGN as presented here.

\subsection{IR anisotropy estimation is Seyfert galaxies}

\citet{Buc06} used a sample of local Seyfert galaxies and compared the average 5--35\,$\micron$ SED of type 1s and type 2s, scaled to their respective 8.4\,GHz emission. However since the sources have been selected according to a flux limit at 12\,$\micron$, the sample cannot be considered isotropic. \citet{Buc06} report generally higher fluxes for type Seyfert 1 AGN as compared to Seyfert 2s. The anisotropy decreases from a factor of about 8 to $\sim$2.5 from 5 to 8\,$\micron$. This is significantly larger than what we find for our isotropically selected radio-loud sample. At longer wavelengths the Seyfert 1/Seyfert 2 ratio of \citet{Buc06} flattens to about a factor of 2--3 (with no convergence to unity as expected at long wavelengths) which is, again, larger than what we found. 

The discrepancy between our results and \citet{Buc06} may be either due to (1) the different selection criteria chosen, (2) a difference between low and high luminosity AGN, or (3) a difference between radio-quiet and radio-loud AGN. While the data in \citet{Buc06} has not been corrected for host galaxy, the contribution by starformation should not be significant given the lack of PAH features in the average spectrum. If a host-correction were applied, it would predominantly affect type 2 AGN, thus making the anisotropy even larger. Furthermore, if the mid-IR selection had any effect, then the sample would miss out AGN at highest obscuration, so that the real type 1/type 2 anisotropy would again be larger. In conclusion, possible selection effects and host contamination in the \citet{Buc06} study would tend to result in an underestimated anisotropy, making the difference to our findings even stronger.

It is well possible that the difference in anisotropy between our high-luminosity radio-loud sample and the low-luminosity radio-quiet sample is real. This would imply that either luminosity or radio power are the drivers for the observed characteristics. Radio jets are highly collimated so that any influence of the jet can be expected perpendicular to the torus plane and just in a very small solid angle. It is, therefore, more reasonable to assume that the higher AGN luminosity would cause the lower anisotropy than the jet. The classical receding torus picture changes the opening angle of the torus for higher luminosity \citep[e.g.][]{Law91,Sim05}. This mainly affects the relative number of type 1s and type 2s in a sample. The relatively high fraction of about 50\% unobscured AGN in our sample would support this scenario. However, to change the anisotropy between both types, it would be necessary to also change the obscuration properties (i.e. type 2s must on average look more like type 1s). Such a scenario of ``radiation-limited obscuration'' has been proposed by \citet{Hon07} and supported by observations of \citet{Tre09}. In this case about the same effect is expected for radio-quiet and radio-loud AGN.

\section{Summary and conclusions}\label{sec:summary}

We use the sample of quasars and radio galaxies at $z\sim1.2$ recently presented in \citet{Lei10}. Since the sample was selected isotropically, it should cover all torus inclination angles (weighted by solid angle). For these objects an infrared 1--17\,$\micron$ restframe SED has been constructed. Average SEDs were calculated for the quasar (= type 1 AGN) and radio galaxy (= type 2 AGN) samples, respectively, in order to study the intrinsic anisotropy of the IR emission of the dust torus. 

It is shown that the ratio between type 1 and type 2 AGN in our parameter space of very luminous radio galaxies, the value gradually decreases from 20 to 2 at wavelengths 2 to 8\,$\micron$. Within the 10\,$\micron$ silicate feature the ratio raises slightly. At longer wavelength the mid-IR emission becomes more isotropic. The intrinsic ratio between our type 1 and type 2 AGN is $1.44\pm0.17$ at 15\,$\micron$. When using IR-selected flux-limited samples this anisotropy has to be taken into account.

By analyzing the silicate feature in the sample averages we find the well-established ``shift'' of the central peak of the silicate emission feature with respect to the center of the absorption feature. The resulting central wavelengths at $9.6\pm0.1\,\micron$ for the absorption and $10.6\pm0.1\,\micron$ for the emission feature are in agreement with previous reports \citep[e.g.][]{Stu05,Shi06}.

We discussed our results in the frame of other anisotropy estimators. Our findings are consistent with upper limits derived from the X-ray/mid-IR correlation of local Seyfert galaxies \citep{Gan09}. Some discrepancy exists with respect to a similar study of \citet{Buc06} for nearby Seyferts. If real it would imply that nearby, radio-quiet lower-luminosity AGN show a higher degree of anisotropy in the IR than higher luminosity, radio-loud sources. This may be explained by a receding torus model with luminosity-dependent obscuration. We also show that the overall relatively small degree of anisotropy is consistent with the torus being clumpy rather than smooth. Our clumpy torus model reproduces the observed type 1/type 2 ratio reasonably well.

\acknowledgments

We would like to thank our referee Prof. Andy Lawrence for helpful and constructive comments which significantly improved the paper, as well as Poshak Gandhi who also commented on this manuscript. The paper was made possible by Deutsche Forschungsgemeinschaft (DFG) in the framework of a research fellowship (``Auslandsstipendium'') for SH. MH is supported by the Nordrhein-Westf\"alische Akademie der Wissenschaften und der K\"unste. This work is based on observations made with the Spitzer Space Telescope, which is operated by the Jet Propulsion Laboratory, California Institute of Technology under a contract with NASA.

{\it Facility:} \facility{Spitzer}

\end{document}